\documentclass[superscriptaddress,reprint,prl,amsmath,amssymb,twocolumn]{revtex4-1}
\usepackage{graphicx}
\usepackage{dcolumn}% Align table columns on decimal point
\usepackage{bm}% bold math
\usepackage{color}
\usepackage{float}
\usepackage{booktabs}
\usepackage{listings}
\usepackage{multirow}
\usepackage[ruled,linesnumbered]{algorithm2e}
\usepackage{algpseudocode}
\usepackage{lineno}
\usepackage{caption}
\begin{document}

% \preprint{APS/123-QED}

\title{Tracking metastable phases by complex Lee-Yang zeros}

\author{Yi-Hua Dong}
\affiliation{State Key Laboratory for Artificial Microstructure and Mesoscopic Physics, Frontier Science Center for Nano-optoelectronics and School of Physics, Peking University, Beijing 100871, P. R. China}
% \author{Fang-chen Wang}
% \affiliation{State Key Laboratory for Artificial Microstructure and Mesoscopic Physics, Frontier Science Center for Nano-optoelectronics and School of Physics, Peking University, Beijing 100871, P. R. China}
\author{Ling Liu}
\affiliation{State Key Laboratory for Artificial Microstructure and Mesoscopic Physics, Frontier Science Center for Nano-optoelectronics and School of Physics, Peking University, Beijing 100871, P. R. China}
\author{Fang-Cheng Wang}
\affiliation{State Key Laboratory for Artificial Microstructure and Mesoscopic Physics, Frontier Science Center for Nano-optoelectronics and School of Physics, Peking University, Beijing 100871, P. R. China}
\author{Qi-Jun Ye} 
\email{qjye@pku.edu.cn}
\affiliation{Interdisciplinary Institute of Light-Element Quantum Materials, Research Center for Light-Element Advanced Materials, and Collaborative Innovation Center of Quantum Matter, Peking University, Beijing 100871, P. R. China}
\affiliation{State Key Laboratory for Artificial Microstructure and Mesoscopic Physics, Frontier Science Center for Nano-optoelectronics and School of Physics, Peking University, Beijing 100871, P. R. China}
\author{Xin-Zheng Li}
\email{xzli@pku.edu.cn}
\affiliation{Interdisciplinary Institute of Light-Element Quantum Materials, Research Center for Light-Element Advanced Materials, and Collaborative Innovation Center of Quantum Matter, Peking University, Beijing 100871, P. R. China}
\affiliation{State Key Laboratory for Artificial Microstructure and Mesoscopic Physics, Frontier Science Center for Nano-optoelectronics and School of Physics, Peking University, Beijing 100871, P. R. China}
\affiliation{Peking University Yangtze Delta Institute of Optoelectronics, Nantong, Jiangsu 226010, P. R. China}

% \linenumbers
\begin{abstract}
Metastable phases (MPs) are energetically unfavorable states typically suppressed in equilibrium phase diagrams.
Rather than remaining ``hidden'', we show that they exist in the complex plane of thermal fields, as regions delineated by Lee-Yang zeros (LYZs).
We demonstrate this numerically in a toy model with a tunable density of states featuring three Gaussian peaks and in a more realistic periodically driven system. 
In both cases, as artificial parameters or drive amplitudes increase, the LYZs bounding the MP approach the real axis and split into separated branches, signaling the emergence and stabilization of the MP within the enlarged gap between two adjacent stable phases.
In the driven system, the imaginary part of LYZs correlates with drive strength, linking Lee-Yang theory to terahertz matter manipulation. 
These findings provide a scheme to describe MPs in phase diagram analysis.
By viewing periodic drives as complex thermal fields, it also offers a new perspective for understanding and engineering non-equilibrium collective states.

\end{abstract}

%\pacs{05.70.Fh, 64.60.Bd, 05.40.Jc, 64.70.Rh} % Phase transitions: general studies; Lee-Yang zeros; Langevin equations; Metastable phases

% Show writing date (optional). Remove or change if you don't want the date printed.
\date{\today}

\maketitle

\textit{Introduction}---Recent years have witnessed remarkable advances in matter manipulation and engineering towards metastable phases (MPs)~\cite{qi2009,Li2019,Nova2019,Ilyas2024}.
As collective states sharing identical chemical composition, MPs exhibit a richer diversity of structures and can host exotic properties that are often absent or less pronounced in established stable phases, such as supercooled liquids, glasses, martensite, and polymorph~\cite{Debenedetti1997,Debenedetti2001,ediger2000,poole1992,nishiyama2012martensitic,greninger1949mechanism}.
However, unlike stable phases, which are reliably mapped by equilibrium phase diagrams (EPDs), the exploration of MPs continues to rely heavily on chemical intuitions drawn from analogous compounds or computationally exhaustive structure predictions.
One significant challenge is that MPs inherently correlate with dissipations, fluctuating orders, and finite lifetimes.
These features arise from realistic stimuli, rendering stable phases and conventional theories of phase transition more the exception than the rule~\cite{Langer1974,Skripov1992}.
Consequently, key hypotheses establishing equilibrium thermodynamics often become incongruous or even misleading when applied to MPs~\cite{palmer1982}.
For example, the thermodynamic limit exaggerates the free energy differences and overlooks MPs entirely.
Theoretical efforts, such as nucleation theory and spinodal decomposition~\cite{cahn1958,cahn1961,alert2016,sleutel2014}, primarily address kinetics and energetics of how metastable local minima decay towards global minima, paying limited attention to MPs' intrinsic existence as reproducible states within certain intensive variables.
Meanwhile, molecular simulation methods, represented by kinetic Monte Carlo and quenched molecular dynamic (MD), can capture the non-equilibrium dynamics and finite-size effects relevant to MPs~\cite{emo2014,jund1997,Sosso20167078,yang1988}.
However, these approaches depend strongly on prior knowledge of relevant pathways or initial conditions, limiting their ability to probe MPs in a proactive and systematic manner.
To uncover the ``hidden iceberg'' beneath stable phases, phase diagrams native to MPs are highly desirable.
Intuitively, new boundaries demarcating MPs are superimposed onto the EPD, as illustrated by spinodal line or dynamical criteria in studies of the diamond-graphite transition and water's liquid-liquid phase transition~\cite{bundy1963,kim2020,luo2022b}.
Nevertheless, this strategy evokes a fundamental dilemma: a single point in the phase diagram may correspond to multiple distinct states, necessitating additional degrees of freedom (DOFs) to distinguish them.
A key opportunity lies in the partition function, which retains full state statistics including MP signatures.
Lee-Yang theory converts this information into a readable phase diagram structure via Lee-Yang zeros (LYZs), i.e., a complex phase diagram~\cite{Yang1952,Lee1952,Ouyang2024}, promisingly placing stable phases and MPs on the same map.
Within this framework, the EPD is a special case where phase boundaries are set by LYZs pinching the real axis in the thermodynamic limit; away from the real axis, additional LYZ structures could contain information relevant to MPs.
In fact, although born for equilibrium thermodynamics, LYZ analysis has been extended to dynamical phase transition, non-Hermitian physics, and supercritical phenomenon, where the imaginary thermal fields serve as the pursued extra DOFs~\cite{bena2005,wei2012,heyl2013,brandner2017,ye2023a,gao2024,Ouyang2024,gu2026a,guo2026,lv2026,liu2023c,lu2025,meng2025,meng2026}.

\begin{figure*}[htbp]
\centering
\includegraphics[width=\linewidth]{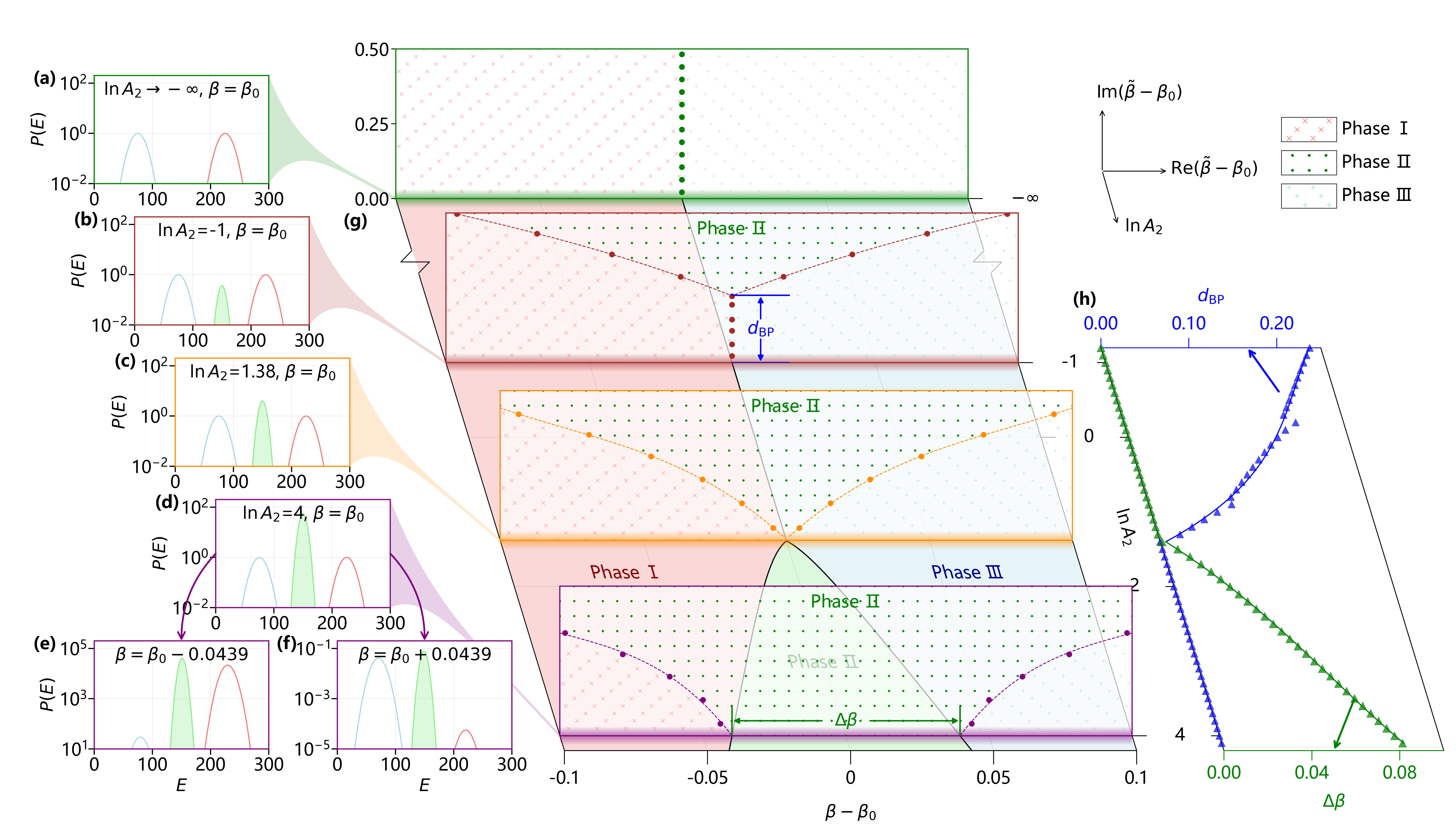}
\caption{The three-Gaussian-peak model with (a)-(d) the referenced DOS distribution $P(E,\beta=\beta_0)$ for varying $A_2$ at $\ln A_2 = -\infty$, $-1$, $1.38$, and $4$, respectively. 
The other parameters are fixed, as $A_1=A_3=1$, $\mu_{k} = 300-75\times k$, $\sigma_{1,3}=10$ and $\sigma_2=5$. 
The light red, green, and blue peaks represent phase I, II, III, respectively. 
For each DOS, it can be reweighted to the profile of coexistence of two phases at certain $T$, such as $\beta=\beta_0\mp 0.0439$ in (e) and (f) from (d), where the phase transitions I-II and II-III occur, respectively. 
(g) the complex phase diagram: the horizontal plane is spanned by real axes of temperature $\beta-\beta_0$ and peak parameter $\ln A_2$, showing the EPD, and the vertical planes spanned by complex temperature axes of $\beta-\beta_0$ and $\text{Im}(\tilde{\beta})$, displaying the profile of LYZs at each $\ln A_2$. 
Each vertical plane is colored corresponding to the DOS in (a)-(d). 
(h) the characteristic of the LYZs as the distance of branching point from the real plane $d_{\text{BP}}$, and the gap between two branches $\Delta \beta$, at varying values of $\ln A_2$. 
(g) and (h) share the $\ln A_2$ axis.
}
\label{fig:p1}
\end{figure*}

Here, we investigate MPs by analyzing LYZs in the complex phase diagram.
Our key finding is that the LYZ structure reveals both thermodynamically stable phases and MPs, with metastable regions remaining clearly delineated by LYZs deep in the complex plane---even before they stabilize on the real axis.
Using a simplified toy model featuring a tunable three-peak density of state (DOS), we demonstrate that specific LYZ loci in the complex plane directly outline the MP boundaries.
As parameters are tuned to enhance the MP, we track these LYZ trajectories and observe their progressive approach to the real axis and subsequent splitting into separate branches, signaling the emergence and stabilization of an MP intervening between two originally successive stable phases, respectively.
We then examine a more realistic oscillatory-field-driving system where a similar process is reproduced but now by tuning the drive strength.
We find that the imaginary parts of the LYZs correlate with the drive strength and the magnitude of DOS peak representing the MP, linking Lee-Yang theory to matter manipulation via terahertz technology.
Thus, by viewing periodic drives as complex thermal fields, one can understand and engineer non-equilibrium collective states in a proactive manner.
\textit{Three-Gaussian-peak model}---The EPD arises from the $T$-dependent reweighting of the thermally populated DOS $P(E, \beta)$ by the Boltzmann factor $e^{-\beta E}$, as different peaks prevail and the corresponding phases hold their own $\beta$-regimes.
Without losing generality, we study a minimal model where the referenced $P(E, \beta=\beta_0)$ consists of three Gaussian peaks, as
\begin{equation}
	P(E, \beta=\beta_0) = \sum_{i=1}^{3} A_i \exp\left( -\frac{(E - \mu_i)^2}{2\sigma_i^2} \right),
\end{equation}
where $\mu_1>\mu_2>\mu_3$ depicts the sequence of three possible phases (I, II, and III, respectively), and $A_i$ is an artificial parameter tuning the strength of the $i$th peak.
We start by looking at the EPDs, which can be obtained directly from these $P(E, \beta=\beta_0)$s, as shown in Fig.~\ref{fig:p1}(a-f), corresponding just to the real axes in Fig.~\ref{fig:p1}(g).
In the limit $A_2=0$, this model presents a conventional two-phase behavior: phase I and III coexist at $\beta=\beta_0$ [as indicated by the approaching $\mu_1$- and $\mu_3$-peaks in Fig.~\ref{fig:p1}(a)] and dominate the $\beta<\beta_0$ and $\beta>\beta_0$ regions in the EPD, respectively.
For moderate $A_2$, this two-phase picture is retained and the intermediate phase II is suppressed in the EPD [as the lowest green peak in Fig.~\ref{fig:p1}(b)].
Upon increasing $A_2$ with $A_{1,3}$ fixed, the EPD undergoes a qualitative change to a three-phase structure: the previously metastable phase II gradually gains stability, first emerging at the triple-point coexistence [Fig.~\ref{fig:p1}(c)], beyond which its thermodynamically stable regime grows and separates phases I and III [Fig.~\ref{fig:p1}(d)].
In this region, for each $A_2$, e.g. $\ln A_2=4$ in Fig.~\ref{fig:p1}(d), the I-II and II-III phase transitions occur as coexistence between neighboring phases at decreased and increased $\beta$ [Fig.~\ref{fig:p1}(e) and (f)], respectively.
Only after phase II becomes stable does the EPD abruptly show signatures of it. 
Before that, although the phase-II peak evolves continuously with $A_2$, phase II remains completely invisible in the two-phase EPD regime [the real plane above the orange thick solid lines in Fig.~\ref{fig:p1}(g)].
In contrast, the complex LYZs can faithfully track the MP through their nontrivial structures, even when phase II lies far from thermodynamic stability.
To show this, we illustrate the profiles of LYZs in the vertical $\tilde{\beta}$-plane for $A_2$ corresponding to Fig.~\ref{fig:p1}(a)-(d).
Herein, the LYZs are derived analytically by extending $\beta$ to the complex variable $\tilde{\beta}=\text{Re}(\tilde{\beta}) + i \text{Im}(\tilde{\beta})$ and solving for the roots of $Z(\tilde{\beta})=0$, where $Z(\tilde{\beta})$ is the partition function given by 
\begin{equation}
	Z(\tilde{\beta}) = \sum_E \rho(E) e^{-\tilde{\beta} E} \sim \sum_E P(E, \beta_0) e^{-(\tilde{\beta} - \beta_0) E}.
\end{equation}
Please see Supplemental Material (SM) for details~\cite{SM}.
In the following, we use $\beta=\text{Re}(\tilde{\beta})$ and $\tilde{\beta}$ to denote the real and complex thermal fields, respectively.
In the limiting case $\ln A_2\to-\infty$, the LYZs form a single continuous boundary that divides the complex plane into two domains [upper green vertical plane in Fig.~\ref{fig:p1}(g)].
The left and right domains, each encompassing half of the real axis, naturally correspond to the two stable phases, I and III, respectively. 
Upon increasing $A_2$, the boundary first branches and a new domain emerges at large $\text{Im}(\tilde{\beta})$-value [red vertical plane in Fig.~\ref{fig:p1}(g)].
As the branching point gradually approaches the real axis [yellow vertical plane in Fig.~\ref{fig:p1}(g)], a triple point between phases I, II, and III appears.
After this triple point, the boundary splits into two separated ones [purple vertical panel in Fig.~\ref{fig:p1}(g)], signaling an abrupt change from an I-III phase sequence to an I-II-III one.
Crucially, this continuous evolution demonstrates that the regions beyond the branching point at low $A_2$ and the intermediate region at high $A_2$ belong to the same phase, i.e., phase II, which leaves no direct signature on the real-axis EPD before stabilization.
Please see the video in the SM for the evolution described above~\cite{SM}.
Based on these, the complex phase diagram---the full LYZ distribution on the vertical planes in Fig.~\ref{fig:p1}(g)---not only captures all visible features of the EPD, but also reveals hidden MPs.
It tracks the entire stabilization process of an MP, from deep in the complex plane prior to its emergence on the real axis.
We quantify the stability of an MP by $d_{\text{BP}}$, the distance from the real axis to the nearest complex branching point: larger $d_{\text{BP}}$ indicates lower stability.
Figure~\ref{fig:p1}(h) compares the evolution of $d_{\text{BP}}$ and $\Delta\beta$ with $\ln A_2$, where $\Delta\beta$ (the gap between phases I and III) measures the stability window of phase II as a thermal ground state.
Before the triple point, $\Delta\beta =0$, so phase II is not a thermal ground state; after this, $\Delta\beta$ increases as phase II stabilizes.
In parallel, $d_{\text{BP}}$ is finite in the metastable regime, decreases monotonically to zero at the triple point, and remains zero thereafter.
In the following, we explore whether an oscillatory field can access such ``hidden'' MP and tune its stability through an experimentally controllable drive strength.

\begin{figure*}[htbp]
\centering
\includegraphics[width=\linewidth]{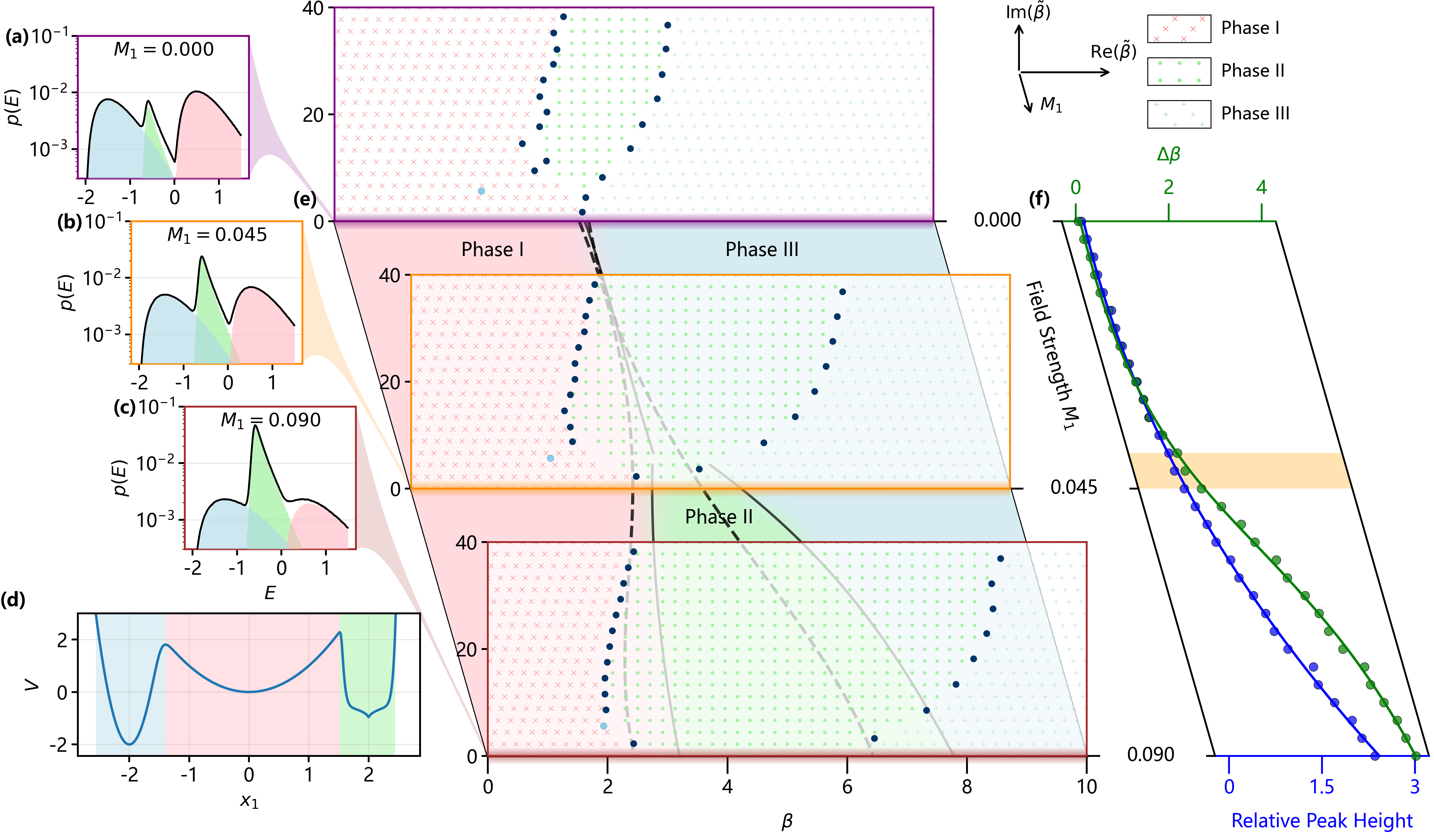}
\caption{The energy probability distribution $P(E, \beta=4)$ under varying field intensities: (a) $M_1=0$, (b) a weak external field with $M_1 = 0.045$, and (c) a strong external field with $M_1 = 0.090$. (d) The potential-energy profile $V(\mathbf{r})$ along the principal direction $x_1$ at $r^\prime =0$. (e) The $M_1$--$\tilde{\beta}$ phase diagram. In the $M_1$--$\text{Re}(\tilde{\beta})$ plane, solid lines are obtained from MD simulations, while the dashed lines label the trajectories of the two closest LYZs. In the $\text{Re}(\tilde{\beta})$--$\text{Im}(\tilde{\beta})$ plane, the calculated LYZ points are plotted. Phases~I--III and their energetic peaks and domains in the complex plane are labeled with light red, green, and blue shading, respectively. (f) $\Delta \beta$ and the relative peak height during tuning, with the latter defined as the $p(E)$ area ratio of phase II to the others. The shadow (e) and (f) share the $M_1$-axis.
}
\label{fig:p2}
\end{figure*}

\textit{Oscillatory-field-driving system}---Matter manipulation and structure engineering with terahertz pulses were reported in recent studies, demonstrating the capability to selectively excite specific vibrational modes and thereby access MPs that are otherwise kinetically inaccessible under equilibrium conditions~\cite{Nova2019,Li2019,Ilyas2024}.
Building on the three-Gaussian-peak toy model, we construct a periodically driven counterpart that replaces artificial peak height tuning with experimentally controllable external fields.
Specifically, we perform Langevin MD on a designed three-well potential (see Fig.~\ref{fig:p2}(d) for the shape and Methods for details) and drive the principal mode with two frequencies, scanning the drive amplitude $M_1$ (with fixed $M_2=2M_1$) to tune the relative heights of the three peaks.
This setup is designed not only to realize the toy-model physics under experimentally tunable driving, but also to link the control parameter $M_1$ to LYZ-based complex phase-diagram analysis.
Model construction and implementation details are provided in Methods.

Our MD simulations indeed show the emergence and stabilization of the hidden phase.
When no external field is applied, as shown in Fig.~\ref{fig:p2}(a), phase~II is bypassed, as phase~I and phase~III already coexists at the reference temperature $\beta_0=4$.
This mirrors the scenario of Fig.~\ref{fig:p1}(b), where phase II is metastable.
With driving, i.e., by tuning $M_1$ and $M_2$, the states corresponding to phases~I and~III are excited and become less favorable, rendering the minimally affected phase~II more accessible.
Correspondingly, as shown in Fig.~\ref{fig:p2}(b) and (c), the central peak associated with phase~II gradually stands out and eventually prevails in $P(E, \beta)$ as $M_1$ increases at $\beta_0 = 4$.
Beyond $M_1\sim 0.045$, the system experiences two phase transitions: phase~I-to-II at $\beta_{\text{c}1} < \beta_0$ and phase~II-to-III at $\beta_{\text{c}2} > \beta_0$.

To gain insights into how this MP evolves in the complex phase diagram, we analyze the LYZ profiles during this process when the drive is enhanced.
The LYZs are computed by discretizing the sampled $P(E, \beta)$ with a suitable $\Delta E$ and decomposing the partition function into a polynomial in $e^{-\beta \Delta E}$, following Ref.~\cite{Liu2025}.
The results are shown in Fig.~\ref{fig:p2}(e).
In the horizontal plane of $\text{Re}(\tilde{\beta})$ and $M_1$, which corresponds to the evolution of the thermal ground states at varying $M_1$, the solid lines denote the phase boundaries determined from the MD simulations, and the dashed lines indicate the two closest LYZs featuring phase transitions.
The overall trends of the MD solid boundaries and LYZ dashed trajectories are consistent, while quantitative numerical deviations remain and are analyzed in the Sec.~II C of SM~\cite{SM}.
In the vertical planes of $\text{Re}(\tilde{\beta})$ and $\text{Im}(\tilde{\beta})$, we plot the discretized LYZ points.

Overall, Fig.~\ref{fig:p2}(e) displays a continuous field-driven reshaping of phase domains across the complex phase diagram.
When there is no external field, the half complex plane is divided into three regions (upper purple vertical plane): phase I and III claim the real axis, relegating metastable phase II to the complex plane.
As $M_1$ increases, the region associated with phase II gradually approaches and eventually touches the real axis, promoting it to become effectively stable over a certain $\beta$ range (orange and red vertical planes).
Meanwhile, both the dominant $\beta$-ranges of phase I and III are narrowed.
LYZ analysis makes this otherwise compressed and hidden information explicit by resolving drive-induced changes as a clear evolution in the complex phase topology.
Accordingly, the scheme proposed in this manuscript---using complex phase diagrams to characterize the MPs---is not only grounded in the Lee-Yang phase transition theory, but also experimentally accessible through approaches such as terahertz manipulation. 
This establishes a concrete connection between theory and experiment.
To put this connection on a quantitative basis, Fig.~\ref{fig:p2}(f) compares numerical indicators of the same metastability evolution.
As $M_1$ increases, both the relative peak height of phase~II in $P(E,\beta)$ and the LYZ quantity $\Delta\beta$ (as defined earlier, the gap between phase~I and III) evolve consistently and track the onset and stabilization of the MP.
Additional comparisons, including the MD--LYZ deviation $\delta \beta_{c}=\beta^{\text{MD}}_c-\beta^{\text{LYZ}}_c$ and $d_{\text{BP}}$, are shown in Fig.~S9 of~\cite{SM}.
Together, these results show that field-induced modulation of DOS peak weights is faithfully captured by LYZ evolution and connected to experimentally observable phase behavior, while $d_{\text{BP}}$ quantifies experimental accessibility.
Furthermore, this correspondence suggests an effective interpretation of oscillatory driving in terms of a complex thermal field.
Taking the real axis as the common baseline between the EPD and LYZ complex plane, increasing $M_1$ (or, in the toy model, increasing $\ln A_2$) corresponds to moving $\tilde{\beta}$ away from the real axis toward the branching point $\tilde{\beta}_{\text{BP}}$ and beyond, thereby reproducing the emergence and stabilization of the MP.
From this perspective, the LYZ structure of the complex phase diagram provides a practical route to understand and detect driven MPs.
%We note that this is an effective, model-dependent mapping rather than a strict universal equivalence, and a rigorous statistical derivation is left for future work.
%

\textit{Conclusion}---In this work, we establish that the distribution of LYZs deep in the complex plane encodes signatures of metastability, moving beyond the traditional Lee-Yang paradigm that focuses solely on zeros near the real axis. 
In the resulting complex phase diagram, stable phases and MPs can be understood on equal footing as connected domains that are free of and delineated by LYZs.
From this perspective, MPs occupy well-defined locations in the complex plane of thermal fields, rather than being ``hidden'' before stabilization.
This also suggests that stable phases may be the exceptions rather than the rule: among the many LYZ-delineated regions in complex space, only a special subset ``coincidentally'' claims the real axis.
Guided by the complex phase diagram, one can engineer MPs in a proactive manner, e.g., by predicting and scanning MP candidates via DOS simulations and LYZ analysis.
Meanwhile, one might also rationalize terahertz-driven manipulation in a statistical manner, by interpreting periodic drives as effective complex thermal fields.
We anticipate that this framework will enable practical routes for designing novel metastable materials for advanced electronics and quantum technologies.

\section*{Acknowledgements}
We are supported by the National Natural Science Foundation of China under Grant Nos. 12550005, 12522410, 12234001, 12474215, 62321004, and the National Basic Research Programs of China under Grant Nos. 2021YFA1400500 and 2022YFA1403500. We thank the supercomputer center at Peking University for computational resources.

\section*{Appendix: Model construction and molecular dynamics on oscillatory-field-driving system}
To highlight the essential dynamics of the system under periodic driving, we adopt a mean-field-like description of the equations of motion for the degrees of freedom (DOFs).
In this approach, the original many-body interactions are approximated by effective dissipation terms~\cite{Gardiner2004,Sekimoto1998}, as:
\begin{equation}
m\frac{d^2\mathbf{r}}{dt^2} = -\nabla V(\mathbf{r}) - \gamma \frac{d\mathbf{r}}{dt} + \boldsymbol{\xi}(t) + \mathbf{F}_{\text{ext}}(t)
\end{equation}
where $\mathbf{r} = (x_1,x_2,\cdots,x_n)$, $\gamma$ is the friction coefficient, and $\boldsymbol{\xi}(t)$ is Gaussian white noise satisfying the fluctuation-dissipation theorem, as
\begin{equation}
\langle\xi_i(t)\xi_j(t')\rangle = 2\gamma k_B T \delta_{ij}\delta(t-t').
\end{equation}
The $n$ DOFs are divided into one principal mode $x_1$, which couples strongly with the external driving force
\begin{equation}
\mathbf{F}_{\text{ext}}(t) = \left[M_1\sin\left(\frac{2\pi t}{\tau_1}\right) + M_2\sin\left(\frac{2\pi t}{\tau_2}\right)\right] \hat{\mathbf{x}}_1,
\end{equation}
and the bath ones $\mathbf{r}^\prime = (x_2,\cdots,x_n)$, representing the remaining lattice modes uncoupled to it.
The periods $\tau_1$ and $\tau_2$ are chosen to stimulate resonance along $x_1$.
To produce the desired three-peak $P(E, \beta)$, as shown in Fig.~\ref{fig:p2}(a)-(c) has shown, we design the potential [Fig.~\ref{fig:p2}(d)] in the form
\begin{equation}
V(\mathbf{r}) = -\ln\left[e^{-V_1(x_1,r^\prime)} + e^{-V_2(x_1,r^\prime)} + e^{-V_3(x_1,r^\prime)}\right],
\end{equation}
where:
\begin{equation}
	\begin{aligned}
		V_1(x_1,r^\prime) &= x_1^2 + (r^\prime)^2,\\
        V_2(x_1,r^\prime) &= d^{10} + \frac{1}{2}d^{\frac{1}{2}} - 1,\\
        V_3(x_1,r^\prime) &= 16[(x_1+2)^2 + (r^\prime)^2] - 2,
	\end{aligned}
\end{equation}
and $d = \frac{5}{2}  \sqrt{(x_1-2)^2 + (r^\prime)^2}$.
Here, $V_1$ and $V_3$ are harmonic, with different depths, centering locations, and curvatures, resembling the characteristic phonon modes of solid phases, while $V_2$ is entirely anharmonic and off-resonant.
The peaks described by $V_1$, $V_2$, and $V_3$ in $P(E, \beta)$ are labeled as phases I, II, and III for later analysis.
Based on this model, we performed a series of MD simulations at varying $\beta$ to characterize the equilibrium phases: 
when the thermal fluctuations become comparable to the potential-well depth, particles can escape from the high-$\beta$ stable phases and explore the landscape to locate new free-energy minima.
Beyond these drive-free cases, we also monitored the evolution of steady states under oscillatory driving by performing MD simulations with varying drive amplitude $M_1$ (and $M_2$, with the fixed ratio $M_2=2M_1$).
With $\tau_1=4.44$ and $\tau_2=1.11$, the external fields are chosen to be resonant with particles located in the $V_1$ and $V_3$ wells, respectively.
For computational efficiency, we use a moderate number of bath DOFs ($n=6$), while larger $n$ yields consistent results.
For more details, please see Sec.~I~C of the SM~\cite{SM}.
%

% \section*{Data availability}
% The data that support the findings of this study are included in the article and/or the Supporting Information.
% %
% The raw data and analysis codes are available from the corresponding authors upon reasonable request.

% Bibliography
\bibliography{ref}

% \section*{Acknowledgements}
% We are supported by the National Natural Science Foundation of China under Grant Nos. 12550005, 12522410, 12234001, 12474215, 62321004, and the National Basic Research Programs of China under Grant Nos. 2021YFA1400500 and 2022YFA1403500. We thank the supercomputer center at Peking University for computational resources.

% \section*{Author Contributions}
% Q.J.Y. and X.Z.L. conceived and supervised the project. Y.H.D. performed the calculations and data analysis. L.L. and F.C.W. contributed to the discussion and interpretation of the results. All authors contributed to writing and revising the manuscript.

% \section*{Competing Interests}
% The authors declare no competing interests.

\end{document}